# Scaling relations in equilibrium nonextensive thermostatistics

Sumiyoshi Abe[1] and A. K. Rajagopal[2]

[1]*Institute of Physics, University of Tsukuba, Ibaraki 305-8571, Japan*

[2]*Naval Research Laboratory, Washington, DC 20375-5320, USA*

**Abstract**   The forms of Euler and Gibbs-Duhem relations discussed in thermodynamics of extensive systems are shown to hold also for nonextensive systems with long-range interactions with a novel interpretation of the entities appearing therein. In this way, the principles underlying Tsallis' scaling relations in *equilibrium* nonextensive thermostatistics are clarified.

PACS numbers: 05.70.–a, 05.20. –y, 05.90.+m

Nonextensive systems with long-range interactions reside outside of the realm of ordinary thermostatistics [1-3]. Their macroscopic quantities do not scale linearly with respect to system size and therefore the standard thermodynamic limit cannot naively be employed. It is important to notice that they are "small systems" since their elements always feel the effects of the boundaries through the long-range interactions. It seems that these systems have been regarded for a long time as anomalous ones that are of less importance from the pragmatic viewpoint of thermostatistics. The situation has changed, however. For example, there are active experimental and theoretical studies on non-neutral plasmas [4,5]. Also, development of numerical analysis of self-gravitating systems [6] stimulates investigation of traditional gravothermodynamics in a new perspective [7]. One may also add to these the nanometric systems of great current interest [8-12] in view of smallness of the systems. In particular, Hill and his co-worker [8,9] have developed for small systems a modified version of the conventional thermodynamics, which preserves the Euler relation for the internal energy but violates the Gibbs-Duhem relation. In the present paper, we present a different theory, in which the forms of the Euler as well as Gibbs-Duhem relations are retained but with different interpretations of the terms appearing therein.

Let us imagine a generic model with a long-range interaction. As an example, consider a spin model defined on a $d$-dimensional hypercubic lattice whose Hamiltonian reads $H = -J \sum_{i \neq j} \sigma_i \sigma_j / r_{ij}^\alpha$, where $\sigma_i$ is the spin on the $i$th site, $J$ is a positive coupling constant, $r_{ij}$ is the distance between the $i$th and $j$th sites, and $\alpha \, (> 0)$ is a constant characterizing the property of the interaction range. In the continuum approximation, the internal energy per spin can be estimated to be [13,14] $U/N \propto N^*$, where $N$ is the total number of spins. $N^*$ is a factor given as follows:

$$N^* \sim \begin{cases} \text{const} & (\alpha > d) \\ \ln N & (\alpha = d) \\ N^{1-\alpha/d} & (0 < \alpha < d), \end{cases} \qquad (1)$$

in the large-$N$ limit. This factor highlights the nonextensive feature of the system with $\alpha \leq d$ and, in fact, universally appears in various systems with long-range interactions [15,16]. The asymptotic behavior of $N^*$ in Eq. (1) is determined by the property of the power-law decay of the interaction.

The scaling discussion in Ref. [13] is inspired by the nonextensive internal energy, $U \propto N N^*$. There, the Gibbs free energy of a nonextensive system is assumed to be of the same form as that of an ordinary extensive system and to satisfy the following scaling property:

$$\frac{G}{N N^*} = \frac{U}{N N^*} - \frac{T}{N^*}\frac{S}{N} + \frac{P}{N^*}\frac{V}{N}, \qquad (2)$$

where $T$, $S$, $P$, and $V$ are the temperature, the entropy, the pressure, and the volume, respectively. (The magnetic energy may also be added, if necessary.) This equation means that the entropy (as well as the volume) scales with $N$ and, therefore, is *extensive*, whereas the ordinary intensive quantities, the temperature and the pressure, scale with $N^*$ [13,14,16]. This turns out to be consistent with the Monte Carlo results reported in Ref. [17] (see also Ref. [18]).

In what follows, first we reconsider the Gibbs free energy and its scaling property with the viewpoint of retaining the forms of the Euler and Gibbs-Duhem relations, but with altered interpretations of the entities entering therein. Here, we are concerned with the concept of homogeneity and, therefore, we limit ourselves to the strongly nonextensive case

$$0 < \alpha < d. \tag{3}$$

Let us start our discussion by considering the internal energy, $U(S, V, N)$, of a nonextensive system with a long-range interaction, which scales as

$$U(S, V, N) = u(s, v) N N^*, \tag{4}$$

where $u$ is a function of the entropy and volume per particle, $s = S/N$ and $v = V/N$, respectively. Therefore, we have

$$U(\lambda S, \lambda V, \lambda N) = \lambda^{2-\alpha/d} U(S, V, N). \tag{5}$$

Taking the derivatives of the both sides with respect to $\lambda$, we have

$$\frac{\partial U}{\partial(\lambda S)} S + \frac{\partial U}{\partial(\lambda V)} V + \frac{\partial U}{\partial(\lambda N)} N = (2 - \alpha/d) \lambda^{1-\alpha/d} U. \tag{6}$$

Putting

$$\lambda = \left(\frac{1}{2-\alpha/d}\right)^{\frac{1}{1-\alpha/d}}, \qquad (7)$$

we obtain the Euler relation

$$U = TS - PV + \mu N, \qquad (8)$$

where $T = \partial U / \partial S$, $P = -\partial U / \partial V$, and the chemical potential $\mu = \partial U / \partial N$. Note that in the discussion of the extensive systems, one sets $\lambda$ equal to unity. The choice in Eq. (7) is a reflection of the nature of the problem of long-range interactions. Thus, we obtain our first result that the Euler relation for nonextensive systems is formally equivalent to the ordinary Euler relation [19] for extensive systems. This means that the form of the Gibbs-Duhem relation also remains unchanged. From Eqs. (6) and (8), it follows that

$$S dT - V dP + N d\mu = 0, \qquad (9)$$

which is, in fact, formally equivalent to the ordinary Gibbs-Duhem relation for extensive systems. Eq. (8) yields the thermodynamic potential, $G = \mu N$, and therefore Eq. (2) is justified. Since $G \propto N N^*$, the chemical potential scales with $N^*$, similarly to $T$ and $P$, indicating the necessity of incorporating the fluctuations in the system. Consequently, we obtain the following scaling relations:

$$T = T(S, V, N) = \frac{1}{N^*} T\left(\frac{S}{N}, \frac{V}{N}\right), \tag{10}$$

$$P = P(S, V, N) = \frac{1}{N^*} P\left(\frac{S}{N}, \frac{V}{N}\right), \tag{11}$$

$$\mu = \mu(S, V, N) = \frac{1}{N^*} \mu\left(\frac{S}{N}, \frac{V}{N}\right). \tag{12}$$

These relations should be compared with Eqs. (3.20)-(3.22) in Ref. [19], in which the prefactor, $1/N^*$, is absent. The new Gibbs-Duhem relation presented in Eq. (9) shows that there is a new relationship between the parameters defined by Eqs. (10)-(12). With respect to these variables, we may define other thermodynamic quantities such as specific heat, compressibility and so on with similar definitions of magnetization and susceptibility. This justifies the manipulation employed in Ref. [17], in which Monte Carlo calculations of long-range interacting Ising-like systems are performed.

Finally, let us illustrate the above development by taking $S(U, V)$ to be of the form

$$S = c U^\xi V^\eta \tag{13}$$

with constant $c$, $\xi$, and $\eta$. Then, extensivity of $S$ leads to the following condition on $\xi$ and $\eta$:

$$(2 - \alpha / d)\xi + \eta = 1. \tag{14}$$

Then, we obtain the following expressions for the thermodynamic parameters in Eqs. (10)-(12):

$$\frac{T}{N^*} = \frac{1}{\xi} \frac{U}{N N^*} \left(\frac{S}{N}\right)^{-1}, \tag{15}$$

$$\frac{P}{N^*} = \frac{\eta}{\xi} \frac{U}{N N^*} \left(\frac{V}{N}\right)^{-1}, \tag{16}$$

$$\frac{\mu}{N^*} = \frac{1+\eta}{\xi} \frac{U}{N N^*}. \tag{17}$$

In these forms, the scaling properties of the quantities and variables are manifestly appreciated.

In conclusion, we have established the Euler and Gibbs-Duhem relations for equilibrium nonextensive thermostatistics of the systems with long-range interactions. We have seen how this approach can clarify the principles underlying Tsallis' scaling relations. It is of crucial importance to notice that in the present discussion we are concerned with equilibrium. It is generally considered that the large system-size limit and the long-time limit do not commute with each other in nonextensive systems. If the long-time limit is taken first, one observes nonequilibrium stationary states, which survive for very long periods [20]. Currently, there is an ongoing investigation of understanding physical properties of such states based on a generalization of conventional Boltzmann-Gibbs statistical mechanics, termed nonextensive statistical mechanics [21-23], which employs Tsallis' nonadditive entropy [24]. In this respect, attention should be paid to the recent discussion [25] that Tsallis' scaling relations are valid *only in equilibrium states, and not in nonequilibrium stationary states*.

Note Added. The Euler relation in nonextensive systems has been considered in Ref. [26], without reference to the scaling properties. Also, the authors of Ref. [27] discuss the Gibbs-Duhem relation in terms of Tsallis' nonadditive entropy. Therefore, there are no overlaps between the present work and these ones.

S. A. was supported in part by the Grant-in-Aid for Scientific Research of Japan Society for the Promotion of Science. A. K. R. acknowledges the Office of Naval Research for partial support.